\documentclass[amsmath,amssymb,nofootinbib,superscriptaddress,showkeys]{revtex4-2}
\usepackage{graphicx}
\usepackage{dcolumn}
\usepackage{bm}
\usepackage{subfigure}
 \usepackage{multirow} 
 \usepackage{epstopdf}
\usepackage[colorlinks=true, citecolor=blue, urlcolor = blue, linkcolor= red, bookmarks=true]{hyperref}
\usepackage{orcidlink}

\begin{document}

\title{Probing Lorentz Symmetry Violation through Lensing Observables of Rotating Black Holes}

\author{Arun Kumar \texorpdfstring{\href{https://orcid.org/0000-0001-8461-5368}{\orcidlink{0000-0001-8461-5368}}{}}}\email{arunbidhan@gmail.com}
\affiliation{Institute for Theoretical Physics and Cosmology, Zhejiang University of Technology, Hangzhou 310023, China}
\author{Shafqat Ul Islam \texorpdfstring{\href{https://orcid.org/0000-0002-8539-5755}{\orcidlink{0000-0002-8539-5755}}{}}} \email{Shafphy@gmail.com}
\affiliation{Astrophysics and Cosmology Research Unit, 
School of Mathematics, Statistics and Computer Science, University of KwaZulu-Natal, Private Bag 54001, Durban 4000, South Africa}
\author{Sushant~G.~Ghosh \texorpdfstring{\href{https://orcid.org/0000-0002-0835-3690}{\orcidlink{0000-0002-0835-3690}}{}}}\email{sghosh2@jmi.ac.in}
\affiliation{Centre for Theoretical Physics, Jamia Millia Islamia, New Delhi 110025, India}
\affiliation{Astrophysics and Cosmology Research Unit, 
School of Mathematics, Statistics and Computer Science, University of KwaZulu-Natal, Private Bag 54001, Durban 4000, South Africa}
%\date{\today}
\begin{abstract} 
We find a Kerr-like black hole solution—a rotating Bumblebee black hole (RBBH) with a Lorentz-violating parameter $\ell$ and examine the strong lensing by it. The parameter $\ell$ changes the event horizon radius and photon sphere, resulting in a different lensing signature compared to the Kerr black hole of general relativity. Using the strong deflection limit formalism, we compute key observables such as the angular positions of relativistic images, their separation, magnification, and time delays for supermassive black holes Sgr~A* and M87*. Our results show that the parameter $\ell$ has a profound influence on these observables, with $\ell > 0$ suppressing and $\ell < 0$ increasing the deflection angle compared to the Kerr case. We compare RBBH observables with those of Kerr black holes, using Sgr A* and M87* as lenses to observe the effect of the Lorentz symmetry-breaking parameter $\ell$. For Sgr A*, the angular position $\theta_\infty$ in $\in~(18.25-33.3)~\mu as$, while for M87* $\in~(13.71-25.02)~\mu as$. The angular separation $s$, for supermassive black holes (SMBHs) SgrA* and M87*, differs significantly, with values ranging $\in~(0.005-0.81)~\mu as$ for Sgr A* and $\in~(0.003-0.6)~\mu as$ for M87*. The relative magnitude $r_{mag}$ $\in~(3.04-8.15)~\mu as$. We also compared the time delays between the relativistic images in the SMBHs and found that RBBH can be quantitatively distinguished from Kerr black holes. Our analysis concludes that, within the 1$\sigma$ region, a significant portion of the parameter space agrees with the EHT results of M87* and Sgr A*. This demonstrates the feasibility of utilizing strong gravitational lensing to identify Lorentz symmetry violations in extreme gravity regimes. Weak lensing analysis and Einstein ring observations provide further constraints, producing an upper bound of $\ell \lesssim \mathcal{O}(10^{-6})$. 
\end{abstract}


\maketitle
\section{Introduction}
One of the most well-known predictions of general relativity (GR) is gravitational lensing, which is the deflection of light beams by a gravitational field \cite{Weinberg:1972kfs}. With its deep insights into the geometry of spacetime, the nature of gravity, and the distribution of matter on different scales, it has become a potent astrophysical tool \cite{Schneider:1992bmb}.  Direct probes of the spacetime structure and constraints on black hole parameters are made possible by powerful gravitational lensing around black holes, which in particular offers a unique window into the strong-field domain of gravity \cite{Bozza:2010xqn,Virbhadra:1999nm}. Also, gravitational lensing is useful for testing GR \cite{2018grav.book.....M,2021gaie.book.....H}. Through gravitational lensing in the vicinity, a black hole would provide another avenue for testing GR. Together with the first image of the black hole captured by the EHT \cite{EventHorizonTelescope:2019dse, EventHorizonTelescope:2022wkp}, gravitational lensing will also become a vital probe to study the isolated, dim black hole. 

The study of gravitational lensing has a long and fascinating history, originating with the discovery of GR itself \cite{Einstein:1936llh}. According to the investigation \cite{1964MNRAS.128..295R, PhysRev.133.B835, Mellier:1998pk, Bartelmann:1999yn, Perlick:2003vg, Perlick:2004tq, Schmidt:2008hc, Perlick:2010zh, Barnacka:2013lfa}, on gravitational lensing has significantly improved our understanding of spacetime. One of the first researchers to apply gravitational lensing principles to Schwarzschild black holes was Darwin \cite{1959RSPSA.249..180D}; Luminet \cite{Luminet:1979nyg} went on to develop these ideas.  He derived a logarithmic approximation for light rays moving near the photon sphere, now known as the strong deflection limit. Atkinson {it et al.} also conducted significant studies on this topic \cite{1965AJ.....70..517A,2018grav.book.....M,Luminet:1979nyg,1987AmJPh..55..428O}. Pioneering theoretical work, primarily by Virbhadra and Ellis \cite{Virbhadra:1999nm,Virbhadra:2002ju}, has focused on the formation and location of rings and the magnification of relativistic images around Schwarzschild black holes.  Later, Frittelli, Kling, and Newman \cite{Frittelli:1999yf} obtained solutions to the exact lens equation in the form of integral expressions.
Subsequently, Bozza \cite{Bozza:2001xd, Bozza:2002zj, Bozza:2007gt, Bozza:2002af} and Tsukamoto \cite{Tsukamoto:2016jzh,Tsukamoto:2016qro} developed analytical techniques for strong-gravitational lensing in static and spherically symmetric spacetimes. Higher-order image positions and magnifications have also been obtained for Schwarzschild black holes \cite{Bozza:2001xd} and generic spherically symmetric spacetimes \cite{Bozza:2002zj} and rotating black holes \cite{Bozza:2002af}.  Torres \cite{Eiroa:2002mk} provided analytical formulas for the positions and magnifications of relativistic images in the context of Reissner-Nordström black hole lensing. Perlick \cite{Perlick:2004tq} has also explored the exact gravitational lens equation in spherically symmetric and static spacetimes.
For a detailed discussion on higher-order images and related topics, refer to Bozza \cite{Bozza:2010xqn}. Since then, many studies have focused on gravitational lensing beyond the weak deflection approximation, especially to distinguish different theories of gravity  \cite{Claudel:2000yi, Hasse:2001by,Perlick:2004tq, Iyer:2006cn, Virbhadra:2007kw,Bisnovatyi-Kogan:2008lga, Mukherjee:2009zz, Tarasenko:2010zz, Eiroa:2010wm, Wei:2011nj, Li:2014hvf, Alhamzawi:2016bql, Tsukamoto:2016qro, Aldi:2016ntn, Dai:2018fxc, Aratore:2021usi, Kuang:2022ojj, Aratore:2024ilt}. These methods are also widely accepted in the scientific community and are often used to study strong gravitational lensing by black holes in modified gravity theories \cite{Ghosh:2020spb, Whisker:2004gq, Abbas:2021whh, Eiroa:2005ag, Gyulchev:2006zg, Ghosh:2010uw, Gyulchev:2012ty, Molla:2023hog, Grespan:2023cpa, Kumar:2022fqo, KumarWalia:2022ddq, Kumar:2021cyl, Lu:2021htd, Ali:2021psk, Hsieh:2021scb}. The strong gravitational field remains a major area of research, with recent studies examining lensing effects from various black holes \cite{Chen:2009eu,Sarkar:2006ry,Javed:2019qyg,Shaikh:2019itn} and modifications to the Schwarzschild geometry \cite{Eiroa:2010wm,Ovgun:2019wej,Panpanich:2019mll,Bronnikov:2018nub,Shaikh:2018oul, Lu:2021htd,Babar:2021nst}, including those in higher curvature gravity \cite{Kumar:2020sag,Islam:2020xmy,Narzilloev:2021jtg}. Particularly, the strong gravitational lensing near black holes and compact objects reveals phenomena like shadows, photon rings, and relativistic images \cite{Synge:1966okc,Gralla:2019xty,1959RSPSA.249..180D,Cunha:2018acu, Bozza:2010xqn,Bozza:2001xd,Bozza:2002zj,Bozza:2002af,Gralla:2020yvo, Johnson:2019ljv,Gralla:2020srx, Gralla:2019drh, Wielgus:2021peu, Broderick:2021ohx,Ayzenberg:2022twz,Guerrero:2022msp, Bisnovatyi-Kogan:2022ujt,Tsupko:2022kwi,Eichhorn:2022oma, Broderick:2023jfl, Kocherlakota:2023qgo, Kocherlakota:2024hyq, Aratore:2024bro}. The studies of strong-field gravitational lensing have also received significant attention within the framework of string theories \cite{Bhadra:2003zs, Sharif:2017ogw, Younesizadeh:2022czv, Molla:2024lpt}.

GR has succesfully passed several observational tests \cite{Will:2014kxa}, and astrophysical black holes in our Universe are forecast to be described by Kerr's metric. Black hole in modified theories of gravity (MTG) are the subject of extensive research, particularly since the accelerated expansion of the Universe, revealed by astronomical observations \cite{SupernovaSearchTeam:1998fmf,SupernovaCosmologyProject:2003dcn,SupernovaCosmologyProject:1998vns,Boomerang:2000efg,Riess:2006fw}, can be explained, at least in part, within the framework of MTG \cite{DeFelice:2010aj}.  Furthermore, there is no concrete proof that black hole candidates are Kerr black holes. GW150914 \cite{LIGOScientific:2016aoc,LIGOScientific:2016lio} and other recent observations do not rule out the potential that these black holes' geometry could deviate greatly from Kerr's metric. 

Strong gravitational lensing by rotating black holes in Bumblebee gravity offers a unique way to explore how the Lorentz violation (LV) parameter affects the lensing phenomena and the structure of spacetime around black holes. The study of strong-field gravitational lensing for slowly rotating Kerr-like black holes in Bumblebee gravity has been conducted by Kuang and Övgün \cite{Kuang:2022xjp} and for perturbed slowly rotating Bumblebee black holes by Mangut et al. \cite{Mangut:2023oxa}. These studies are significant because they explore the potential breaking of Lorentz symmetry in gravity and its implications for strong lensing. They use shadow observables from EHT observations to constrain the LV parameter.
Inspired by this, we aim to investigate the strong-field gravitational lensing by RBBH. Further, considering the supermassive black holes Sgr A* and M$87^*$ as the lens with the aid of, we obtain relativistic images' positions, separation, magnification, and time delay. Our results show that it significantly affects strong gravitational lensing.
We will compute observables like the angular position of relativistic images, the distance between images, and the magnification factors by analyzing the deflection angle of light. Our goal is to find unique signatures that might be used as observational probes of Lorentz symmetry violation close to whirling black holes by contrasting these results with those predicted by GR.

This paper is organized as follows: Section \ref{gdapproach} presents the rotating black hole solution in Bumblebee gravity, highlighting its construction and key properties that distinguish it from the Kerr black hole. Section \ref{sgl} is devoted to the analysis of strong gravitational lensing, where we derive the deflection angle, study the formation of relativistic images, and discuss the corresponding lensing observables. We also examine the implications for supermassive black holes Sgr A* and M87*, modelling them as rotating Bumblebee black holes and constraining the parameters using Event Horizon Telescope observations. In Section \ref{wgl}, we turn to weak gravitational lensing and focus on the angular size of the Einstein ring, emphasising its dependence on the Lorentz-violating parameter. Finally, in Section \ref{conclusion}, we summarise the main results and outline possible future directions.

Throughout this work, we employ natural units, setting $G = c = 1$, while the physical constants are restored when presenting numerical estimates.

\section{Rotating black hole in Bumblebee gravity}\label{gdapproach}
We investigate the Bumblebee model, a widely studied framework of MTG in which a vector field—the Bumblebee field—acquires a nonzero vacuum expectation value, leading to spontaneous Lorentz symmetry breaking (LSB). This extension of GR provides a natural setting to explore the effects of Lorentz violation on gravitational physics. A fundamental component of contemporary physics, Lorentz symmetry underlies both the Standard Model and GR. Nonetheless, several quantum gravity perspectives raise the possibility of Lorentz invariance violations \cite{Kostelecky:1988zi,Mattingly:2005re}, which encourages research on Lorentz-violating extensions of GR. Bumblebee gravity is a popular framework in which a vector field gaining a nonzero vacuum expectation value leads to spontaneous Lorentz symmetry breaking \cite{Bluhm:2004ep}.

On the observational side, gravitational wave detections \cite{LIGOScientific:2016aoc}, the Event Horizon Telescope (EHT) images of M87* and Sgr A* \cite{EventHorizonTelescope:2019dse,EventHorizonTelescope:2022wkp}, and enduring cosmological tensions \cite{Riess:2019cxk,DiValentino:2021izs} have made tests of GR in the strong-field regime more significant. Through gravitational lensing and shadows, black holes offer natural laboratories for examining such aberrations.

In this work, we construct a rotating black hole solution in Bumblebee gravity and investigate its lensing properties. Both strong and weak gravitational lensing are analysed, and the resulting observables are compared with EHT data to constrain the Lorentz-violating parameter. We briefly outline the key theoretical underpinnings and distinguishing features of the Bumblebee gravity model in the discussion that follows. Next, as described in Ref.~\cite{Ding:2019mal}, we show a rotating black hole solution that was developed using this framework. This solution is particularly interesting since it provides insights into how LSB affects the characteristics of astrophysical objects by extending the well-known Kerr black hole solution of GR to the Bumblebee gravity context.  

\begin{figure}[t]
	\begin{centering}
		    \includegraphics[width=0.5 \textwidth]{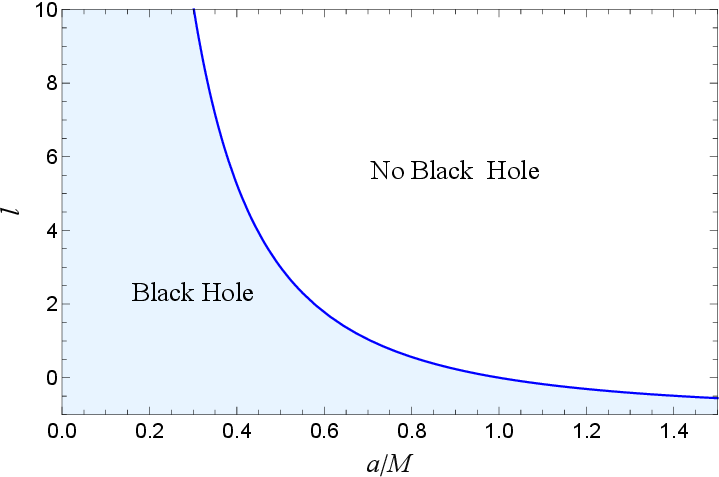}
	\end{centering}
	\caption{Parameter space ($\ell$, $a$) for the existence of RBBH with $M=1$. The black solid curve marks extremal black holes (where the event and Cauchy horizons coincide). The shaded region below the curve represents black holes with distinct horizons (BH region), while the unshaded region (No BH) corresponds to no-horizon solutions.}\label{plot1}		
\end{figure}

The action describing the Bumblebee gravity, reads \cite{Casana:2017jkc,An:2024fzf,Ding:2020kfr,Ding:2019mal}
\begin{equation}\label{action}
    S_{B}=\int d^{4}x \sqrt{-g}[\frac{1}{16\pi G}(R+\gamma B^{\mu}B^{\nu}R_{\mu\nu})-\frac{1}{4}B_{\mu\nu}B^{\mu\nu}-V(B^{\mu})]
\end{equation}
where $R$, $R_{\mu\nu}$, and $\gamma$ are, respectively, the Ricci scalar, Riemann tensor, and coupling constant. The coupling constant $\gamma$ with dimensions $M^{-1}$ defines the non-minimal coupling of gravity with the Bumblebee field ($B_{\mu\nu}=\nabla_{\mu}B_{\nu}-\nabla_{\nu}B_{\mu}$). The Bumblebee potential, $V(B^{\mu})$, which causes the Lorentz and/or CPT (charge, parity, and time), is expressed in the following form
\begin{equation}\label{potential}
    V=V(B^{\mu}B_{\mu}\pm b^{2})
\end{equation}
where $b^{2}$ is positive real constant. The potential (\ref{potential}) yields a non-zero vacuum expectation value, $\left\langle B^{\mu} \right\rangle = b^{\mu}$, corresponding to field $B_{\mu}$. The vector $b^{\mu}$ exhibits a constant value $b_{\mu}b^{\mu}= \mp b^2$ such that $\pm$ signs signify, respectively, the timelike and the spacelike vector $b^{\mu}$. For ensuring the breakdown of $U(1)$ symmetry, the potential must attain minimum at $B_{\mu}B^{\mu}= \mp b^2$ and $V'(b_{\mu}b^{\mu})=0$. The variation of action (\ref{action}) results in the following field equations \cite{Ding:2020kfr,Ding:2019mal} 
\begin{equation}\label{fe}
    G_{\mu\nu}=R_{\mu\nu}-\frac{1}{2}g_{\mu\nu}R = T^{B}_{\mu\nu},
\end{equation} with the Bumblebee energy-momentum tensor $T^{B}_{\mu\nu}$
\begin{eqnarray}\label{em}
T_{\mu \nu }^{B} &=&B_{\mu \alpha }B_{~\nu }^{\alpha }-\frac{1}{4}B_{\alpha
\beta }B^{\alpha \beta }g_{\mu \nu }-Vg_{\mu \nu }+2V^{\prime }B_{\mu
}B_{\nu } +\frac{\gamma }{\kappa }\left[ \frac{1}{2}B^{\alpha }B^{\beta }R_{\alpha
\beta }g_{\mu \nu }-B_{\mu }B^{\alpha }R_{\alpha \nu }-B_{\nu }B^{\alpha
}R_{\alpha \mu }\right.  \notag \\[0.08in]
&&+\frac{1}{2}\nabla _{\alpha }\nabla _{\mu }\left( B^{\alpha }B_{\nu
}\right) +\frac{1}{2}\nabla _{\alpha }\nabla _{\nu }\left( B^{\alpha }B_{\mu
}\right) \left. -\frac{1}{2}\nabla ^{2}\left( B_{\mu }B_{\nu }\right) -\frac{1}{2}%
g_{\mu \nu }\nabla _{\alpha }\nabla _{\beta }\left( B^{\alpha }B^{\beta
}\right) \right] \text{.}
\end{eqnarray} Now, using the conditions, $B_{\mu}=b_{\mu}$, $V=0$ and $V'=0$, the field equations (\ref{fe}) with (\ref{em}) take the following simplified form \cite{Ding:2020kfr,Ding:2019mal,An:2024fzf}
\begin{equation}\label{einstein}
\begin{split}
    R_{\mu\nu}+&\gamma b_{\mu}b^{\alpha}R_{\alpha\nu}+\gamma b_{\nu}b^{\alpha}R_{\alpha\mu}-\frac{\gamma}{2}b^{\alpha}b^{\beta}R_{\alpha\beta}g_{\mu\nu}-\frac{\gamma}{2}\nabla_{\alpha}\nabla_{\mu}(b^{\alpha}b_{\nu})-\frac{\gamma}{2}\nabla_{\alpha}\nabla_{\nu}(b^{\alpha}b_{\mu})+\frac{\gamma}{2}\nabla^{2}(b_{\mu}b_{\nu})=0
\end{split}
\end{equation}
\begin{equation}\label{beq}
    \nabla^{\mu}b_{\mu\nu}=-\frac{\gamma}{\kappa}b^{\mu}R_{\mu\nu}.
\end{equation}
Now, by solving the field equations (\ref{einstein}), we can obtain the Lorentz-violating spherically symmetric  solution in the context of Bumblebee gravity given by the metric \cite{Casana:2017jkc,Ding:2020kfr,Ding:2019mal,An:2024fzf}
\begin{equation}\label{metric}
  ds^{2}=-\left(1-\frac{2M}{r}\right)dT^{2}+\frac{1+\ell}{1-\frac{2M}{r}} dr^{2}+r^{2}d\Omega^{2},
\end{equation} where $M$ is the mass of the black hole and $\ell=\varrho b^2$ is the Bumblebee gravity Lorentz-violating parameter, which differentiates solution (\ref{metric}) from the Schwarzschild black hole solution and takes values in the range shown in Fig. \ref{plot1}. By introducing a transformation such that $t \to \sqrt{1+\ell}~T$, we observe that metric (\ref{metric}) transforms into a Schwarzschild-like solution as:  
\begin{equation}\label{metric1}
\begin{aligned}
 ds^2 = &-(1+\ell)^{-1}\left( 1-\frac{2M}{r} \right) dt^2  + \frac{ dr^2}{ (1+\ell)^{-1}\left( 1-\frac{2M}{r} \right)} + r^2 \left(d\theta ^{2}+\sin ^{2}\theta d\phi ^{2}\right),
\end{aligned}
\end{equation}
The metric (\ref{metric1}) reduces to the Schwarzschild black hole in the absence of a Lorentz symmetry-violating parameter, i.e., $\ell \to 0$. 

\begin{figure}
	\begin{centering}
		\begin{tabular}{ccc}
		    \includegraphics[width=0.5 \textwidth]{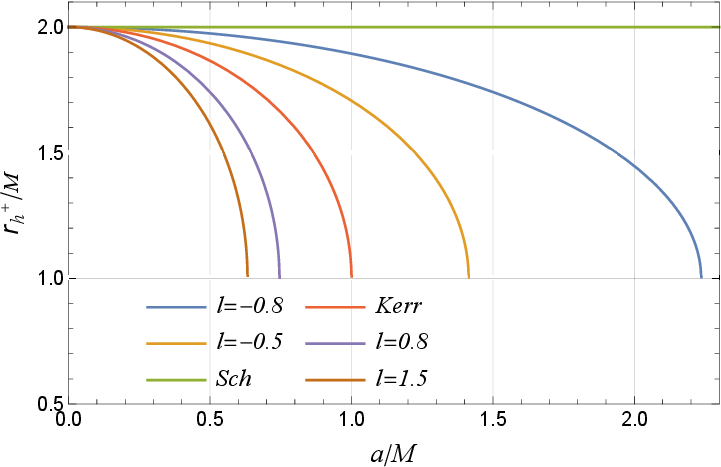}
            \includegraphics[width=0.5 \textwidth]{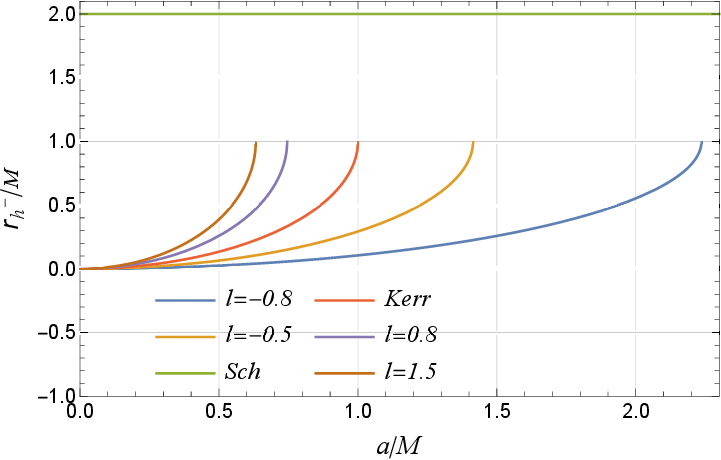}
		    \end{tabular}
	\end{centering}
	\caption{Comparison of the event horizon ($r_h^+$) and Cauchy horizon ($r_h^-$) for RBBH  with a Kerr black hole (red curve) and a Schwarzschild black hole (green line). The Lorentz-violating parameter $\ell$ modified horizon structures: for $\ell > 0$ the horizons contract relative to Kerr ($r_h^+ < {r_h^+}_{\text{Kerr}}$), while for $\ell < 0$ they expand ($r_h^+ > {r_h^+}_{\text{Kerr}}$). The Cauchy horizon $r_h^-$ shows greater sensitivity to $\ell$ than the event horizon.}\label{plot3}		
\end{figure}
Static black holes are useful theoretical concepts in GR, but they are unlikely to exist in natural astrophysical settings. Due to the conservation of angular momentum during gravitational collapse, black holes formed in nature are expected to rotate. Consequently, rotating black holes—characterized by solutions such as the Kerr metric—are regarded as the more physically realistic models. We construct a Kerr-like metric as an axisymmetric extension of the static spacetime given in Eq.(\ref{metric1}). To achieve this rotating geometry, we use a modified version of the Newman–Janis algorithm (NJA) \cite{Azreg-Ainou:2014pra,Azreg-Ainou:2014aqa}. Originally introduced by Newman and Janis in 1965 \citep{Newman:1965tw}, the NJA is a remarkable mathematical technique that allows one to generate rotating (axisymmetric) black hole solutions from static, spherically symmetric seed metrics—without the need to explicitly solve Einstein’s field equations. Over time, this method has been refined and generalized \citep{Azreg-Ainou:2014pra,Azreg-Ainou:2014aqa}, making it applicable to a wider class of spacetimes \cite{Ghosh:2014hea,Jusufi:2019caq,Islam:2024sph}. In our work, we begin with the static, spherically symmetric black hole solution given in Eq.(\ref{metric1}) and, by applying the modified NJA, we obtain its rotating counterpart. The resulting metric is designed to describe rotating black holes within the framework of Bumblebee gravity. To validate this metric, we compare its predictions with observational data from the EHT, which has provided unprecedented insights into the structure of black hole shadows and accretion disks. The rotating counterpart of the black hole solution (\ref{metric}) can be represented by the following metric
\begin{eqnarray}\label{rotating}
ds^2 &=& -\left(1-\frac{2M(r)r }{\Sigma}\right) dt^2+ \frac{\Sigma}{\Delta} dr^2 +\Sigma d\theta^2  +\frac{\mathbb{A}\sin^2\theta~}{\Sigma} d\phi^2 - \frac{4aM(r)r}{\Sigma} \sin^2\theta dtd\phi
\end{eqnarray}
where
\begin{eqnarray}
\Delta &=& r^2+a^2 -2 M(r) r~~~~~
\Sigma = r^2 +a^2\cos^2\theta,~~~~~
M(r) = \frac{M(1+\frac{r\ell}{2M})}{1+\ell},~~~~~~
\mathbb{A} = (r^2+a^2)^2-a^2 \Delta \sin^2\theta. 
\end{eqnarray}
The black hole mass is denoted by \( M \), the LSB parameter by \( \ell \), and the spin parameter by \( a \). For convenience, it is common to work with the dimensionless spin parameter, defined as \( a = a/M \), which normalizes the spin by the mass of the black hole. A non-zero value of \( \ell \) signals a deviation from the standard Kerr solution, arising from LSB. In the special case where \( \ell = 0 \), the solution reduces exactly to the Kerr metric. This recovery of the Kerr solution provides an important consistency check for the theory. Refs.~\cite{Casana:2017jkc,Ding:2019mal} has more detailed discussions of these points.    We shall refer to the rotating metric given in Eq.~(\ref{rotating}) as the \textit{Rotating Bumblebee Black Hole} (RBBH). 

The RBBH in Eq.~(\ref{rotating}) contains two types of singularities, determined by the conditions: \(\Sigma = 0\) and \(\Delta = 0\). The singularity appearing at \(\Sigma = 0\) corresponds to the well-known ring singularity, located in the equatorial plane of the rotating black hole (\ref{rotating}). Its radius is equal to \(a\)-the spin parameter of the black hole, and it represents a fundamental feature of rotating black hole solutions. By contrast, the condition \(\Delta = 0\) identifies the event horizon—the boundary that separates the interior of the black hole from the region outside, beyond which neither light nor matter can escape. The radial position of the event horizon, denoted as $r_{\rm h}$, is obtained by solving equation \(g^{rr} = \Delta = 0\), paralleling the procedure used in Kerr spacetime. The resulting expression provides the horizon radius:
\begin{equation}\label{horizon}
 r_{\pm} = M \pm \sqrt{M^2 - a^2(1 + \ell)} \, .
\end{equation} 
For this solution to be physically meaningful, the condition 
\begin{equation}
|a| \leq \frac{M}{\sqrt{1 + \ell}}
\end{equation}
must be satisfied. This inequality ensures that the event horizon exists and that the black hole solution remains well-defined. The condition also highlights the interplay between the black hole's spin, mass, and the LSB parameter \(\ell\), which collectively influence the structure and properties of the black hole spacetime. Moreover the analysis of (\ref{horizon}) leads us to find out that the geometry can manifest up to two horizons depending upon the values of black hole parameters, $M$, $a$, and $\ell$ such that $-1< \ell< (M^2/a^2-1$) represents the black hole with Cauchy and the event horizons, whereas $\ell=M^2/a^2-1$ represents an extremal black hole. We have illustrated the parameter space diagram in Fig.~\ref{plot1}, where the blue curve corresponds to the extremal black hole solution. The shaded region beneath this curve represents the parameter space for black holes possessing both the Cauchy horizon and the event horizon (denoted as the BH region in Fig.~\ref{plot1}). Above the blue curve, in the region labeled "No BH", no black hole solutions exist. 

The horizon structures of the RBBH are depicted in Fig.~\ref{plot3}. These figures demonstrate that, for a fixed value of the spin parameter $a$, as the parameter $\ell$ decreases, the radius of the Cauchy horizon diminishes, while the radius of the event horizon expands. Furthermore, the event horizon of the RBBH solution exhibits a dependence on the Bumblebee parameter $\ell$. Specifically, when $\ell$ is negative, the event horizon of the RBBH can be larger than that of the Kerr black hole. Conversely, when $\ell$ is positive, the event horizon of the RBBH can be smaller than that of the Kerr black hole. This behavior highlights the significant role of the LSB parameter $\ell$ in modifying the horizon structure compared to the classical Kerr solution.

\paragraph{Surface gravity, Temperature, Area and Angular Velocity}
The physical properties of RBBH can be characterized by the quantities defined at its horizons, which closely resemble those of the Kerr black hole but are modified due to the Lorentz-violating parameter $\ell$. The separation between the inner and outer horizons reads 
The separation between the horizons is
\begin{eqnarray}
	r_+ - r_- = 2 \sqrt{M^2 - a^2(1+\ell)}.
\end{eqnarray}
he separation between the inner and outer horizons determines the causal structure of the spacetime, while the surface gravity $\kappa_\pm$ encodes the strength of the gravitational field at the horizon. 
The surface gravity at the horizons is
\begin{eqnarray}
	\kappa_{\pm} &=& \frac{\Delta'(r_\pm)}{2 \left(r_\pm^2 + a^2\right)}
	= \frac{r_\pm - M}{r_\pm^2 + a^2}
	= \pm \frac{r_+ - r_-}{2 \left(r_\pm^2 + a^2\right)}.
\end{eqnarray}
In particular, for the event horizon
\begin{eqnarray}
	\kappa_{+} = \frac{r_+ - r_-}{2 \left(r_+^2 + a^2\right)}
	= \frac{\sqrt{M^2 - a^2(1+\ell)}}{r_+^2 + a^2}.
\end{eqnarray}
The Hawking temperature $T_H$, proportional to the surface gravity, governs the thermodynamic radiation spectrum of the black hole~\cite{Hawking:1975vcx,Wald:1984rg}.
The Hawking temperature associated with the outer horizon is
\begin{eqnarray}
	T_{H} \;=\; \frac{\kappa_{+}}{2\pi}
	\;=\; \frac{\sqrt{M^2 - a^2(1+\ell)}}{2\pi\,(r_+^2 + a^2)}.
\end{eqnarray}
Similarly, the horizon area $A_\pm$ and entropy $S_\pm$ satisfy the Bekenstein–Hawking area law~\cite{Bekenstein:1973ur,Bardeen:1973gs}, 
The horizon areas and entropies are
\begin{eqnarray}
	A_{\pm} &=& 4 \pi \left(r_\pm^2 + a^2\right),
	\\
	S_{\pm} &=& \frac{A_\pm}{4} \;=\; \pi \left(r_\pm^2 + a^2\right).
\end{eqnarray}
while the angular velocity $\Omega_H^{(\pm)}$ describes the frame-dragging effect at the horizons, as in the Kerr geometry.
The angular velocity of the horizons is
\begin{eqnarray}
	\Omega_{H}^{(\pm)} \;=\; \frac{a}{r_\pm^2 + a^2}.
\end{eqnarray}
These quantities play a fundamental role in black hole thermodynamics and in testing the generalized laws of black hole mechanics in modified gravity ~\cite{Casana:2017jkc,Kumar:2020hgm}.

\section{Strong Gravitational Lensing}\label{sgl}
In the spacetime of an RBBH, we analyze the strong-gravitational deflection of light restricted to the equatorial plane ($\theta = \pi/2$). To keep things simple, we suppose that the source and observer are both asymptotically far from the black hole, meaning that the gravitational field has no effect where they are \cite{Bozza:2002zj,Virbhadra:1999nm,Perlick:2003vg}. In addition to offering analytical tractability, the limitation to equatorial motion also produces insightful results that can be applied to more intricate setups. The metric's reduced version in dimensionless units can be written as
\begin{equation}\label{NSR}
\mathrm{ds^2}=-A(r)dt^2+B(r)\,dr^2 +C(r)d\phi^2-D(r)dt d\phi,
\end{equation}
where
\begin{eqnarray}\label{coeff}
A(r)=-\frac{(2-r)}{(1+\ell) r}\;\;\;\;\;~~~~
B(r)=\frac{r^2}{\Delta},\;\;\;\;\;~~~~~~~
C(r)= \frac{\left(a^2+r^2\right)^2-a^2 \Delta}{r^2},\;\;~~~~~~
D(r)=\frac{2 a (2 + \ell r)}{(1+\ell) r},
\end{eqnarray} 
with $$\Delta= a^2+\frac{r(r-2) }{(1 + \ell)}$$. To analyze photon trajectories, we employ the Hamilton-Jacobi formalism, with the corresponding  Lagrangian for photon as $$ \mathcal{L}=\frac{1}{2}g_{\mu\nu}\dot{r}^{\mu}\dot{r}^{\nu}=0,$$ where the dot denotes differentiation with respect to the affine parameter $\lambda$ along the geodesics. It can be seen that the black hole metric (\ref{NSR}) is characterized by two linearly independent Killing vectors, $\eta^{\mu}{(t)}=\delta^{\mu}t$ and $\eta^{\mu}{(\phi)}=\delta^{\mu}{\phi}$, which correspond to time translation and rotational symmetry \cite{Chandrasekhar:1985kt}. Hence, the path of the photons is influenced by two conserved quantities associated with these Killing vectors: the angular momentum $L$ and the total energy $E$, which are given as follows:
\begin{equation}
E = \frac{\partial \mathcal{L}}{\partial \dot{t}} = -g_{tt} \dot{t} - g_{t\phi} \dot{\phi},~~~~~~~~~~
L = \frac{\partial \mathcal{L}}{\partial \dot{\phi}} = g_{t\phi} \dot{t} + g_{\phi\phi} \dot{\phi}.
\end{equation}
We set $E = 1$ by choosing an appropriate affine parameter and define the impact parameter as $u = L/E$. Using these relations, the equations of motion for photons can be expressed as:
\begin{eqnarray}
\dot{t} &=& \frac{d t}{d\tau}\equiv \frac{r \left(2 a^2 \ell +a^2-a \ell u\right)+2 a^2-2 a u+(\ell+1) r^3}{a^2 (\ell+1) r+r^3-2 r^2},~~~~~~~~~
\dot{\phi} = \frac{d\phi}{d\tau}\equiv  \frac{2 a - 2 u + r (a \ell + u)}{a^2 (\ell+1) r+r^3-2 r^2}.
\end{eqnarray}The above equation, combined with the null condition of the photon ($ds^2=0$), gives the radial geodesic equation as
\begin{eqnarray}
\dot{r} &=& \frac{dr}{d\tau}\equiv \pm 2 \sqrt{\frac{r \left(2 a^2 \ell+a^2-2 a \ell u-u^2\right)+2 a^2-4 a u+(\ell+1) r^3+2 u^2}{4 (\ell+1) r^3}}~~\label{xdot}
 \end{eqnarray}
 Here, the dot denotes differentiation with respect to the affine parameter, while the $+$ and $-$ signs indicate photon motion directed toward and away from the black hole, respectively.
 The radial equation of motion can also be written as 
 \begin{equation}
    \dot{r}^2+V_{\text{eff}}=0
\end{equation}
where $V_{\text{eff}}$ is the effective radial potential, given as \cite{Ghosh:2020spb}:
\begin{eqnarray}\label{effpot}
\frac{V_{\textit{eff}}}{E^2} &=&\frac{r \left(-2 a^2 \ell-a^2+2 a \ell u+u^2\right)-2 a^2+4 a u+(-\ell-1) r^3-2 u^2}{(1+\ell) r^3}.
\end{eqnarray}
\begin{figure*}
	\begin{centering}
		\begin{tabular}{cc}
		    \includegraphics[width=0.5 \textwidth]{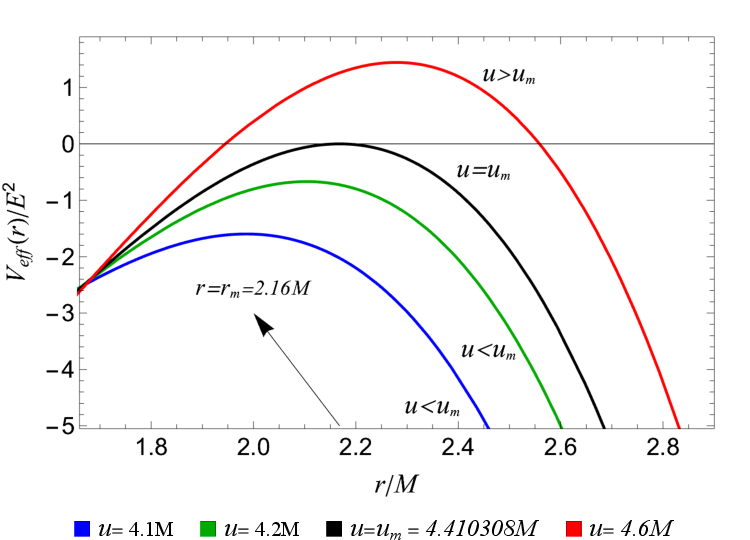}
		  \end{tabular}
	\end{centering}
	\caption{Effective potential $V_{\text{eff}}(r)$ as a function of radial coordinate $r$ for a rotating black hole in Bumblebee gravity with parameters $a = \ell = 0.5$ and energy $E = 1$. The curves represent three distinct regimes: (i) $u > u_m$ (blue, photon deflection), (ii) $u = u_m$ (black, unstable circular photon orbit), and (iii) $u < u_m$ (red, photon capture), where $u_m$ is the critical impact parameter. The photon sphere defines the boundary between scattering and capture paths, and the local maximum at $r = r_m$ corresponds to it.}\label{plot4}		
\end{figure*}  
\begin{figure}
	\begin{centering}
		\begin{tabular}{ccc}
		    \includegraphics[width=0.5 \textwidth]{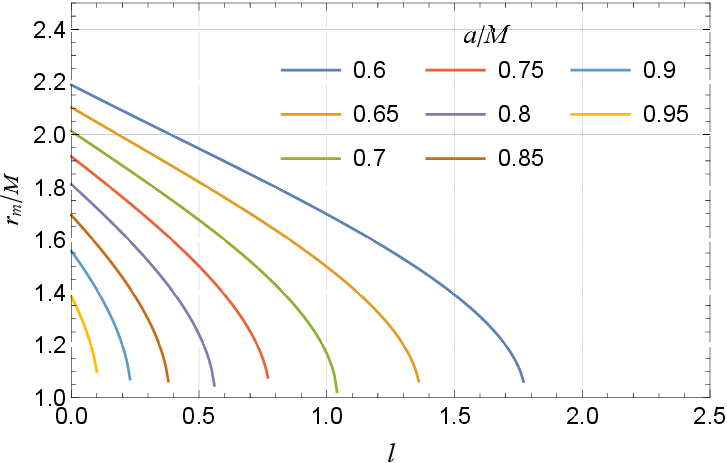}&
		    \includegraphics[width=0.5 \textwidth]{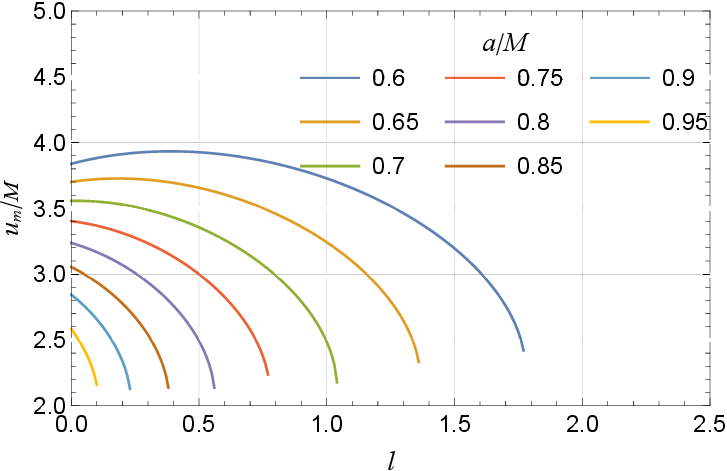}\\
		  \end{tabular}
	\end{centering}
	\caption{ \textbf{Left:} Unstable circular photon orbit radius $r_{\text{m}}/M$ versus Bumblebee parameter $\ell$ for selected spin values $a$. \textbf{Right:} Corresponding critical impact parameter $u_{\text{m}}/M$ as a function of $\ell$. The plots quantify how LSB ($\ell \neq 0$) modifies null geodesics compared to the Kerr solution ($\ell = 0$). For fixed $a$, increasing $\ell$ systematically decreases $r_{\text{m}}$ while producing non-monotonic changes in $u_{\text{m}}$. The $a=0$ case (Schwarzschild limit) shows $\ell$-dependence without spin effects.}\label{plot5}		
\end{figure} 
The effective potential dictates the behaviour of the photon's trajectory around the black hole. A photon can only follow an orbital path if $V_{\textit{eff}}\geq0$, ensuring $\dot{r}^2\geq0$.  The numerical behaviour of the effective potential is depicted in Fig. \ref{plot4}. The local maxima of the effective potential correspond to unstable circular photon orbits, where photons can temporarily orbit the black hole but are highly sensitive to perturbations. Conversely, the local minima of the effective potential indicate the presence of stable circular photon orbits. A light ray originating from infinity and confined to the equatorial plane can be characterized by its impact parameter  $u$. Since this impact parameter and the distance of the closest approach $r_0$ are uniquely connected, defining one defines the other. This indicates that the effective potential meets the condition $V_{\text{eff}} = 0$. At the turning point, when $r = r_0$, the radial component of the photon's velocity momentarily disappears, i.e., $\dot{r}=0$. Solving this equation yields the relation between the impact parameter $u$ and $r_0$ as follows:
\begin{equation} \label{angmom}
L = u(r_0) = \frac{ -a(2 + \ell r_0) \pm (1 + \ell) r_0 \sqrt{ a^2 + \dfrac{(-2 + r_0) r_0}{1 + \ell} } }{ r_0 - 2 }
\end{equation}
Photons that orbit in the same direction as the black hole's rotation (prograde photons) follow a different path compared to those orbiting in the opposite direction (retrograde photons). To ensure counterclockwise winding of light rays, we kept ourselves to the expression of impact parameter with a positive sign in front of the term containing the square root in  Eq.~(\ref{angmom}). 
The deflection angle increases as the impact parameter decreases, eventually exceeding $2\pi$ when the impact parameter reaches a critical value. At this point, photons execute complete orbits, forming what is known as the \textit{photon sphere}. These photon orbits are inherently unstable under small radial perturbations, causing the photons to either fall into the black hole or escape to infinity. The exact locations of photon spheres near a black hole are conventionally determined using the effective potential for photons.
Specifically, the unstable photon spheres correspond to the local maxima of the effective potential $V_{\text{eff}}(r)$, and are identified by the conditions:
\begin{eqnarray}\label{effV}
V_{\text{eff}}=\frac{dV_{{\text{eff}}}}{dr}\Big|_{(r_0=r_m)}=0, ~~~\frac{d^2 V_{{\text{eff}}}}{dr^2}\Big|_{(r_0=r_m)}<0,
\end{eqnarray}
where $r_m$ and $r_0$ represent the photon sphere radius and the light's minimum approach distance. Solving Eq.~(\ref{effV}), the circular unstable photon orbit radius is  given by the following equation: 
\begin{equation}\label{ps}
r \left(-2 r^2+5 r-6\right) +2 a (\ell+1)\left(a- \sqrt{a^2+\frac{(r-2) r}{\ell+1}}\right)\Bigg|_{(r=r_m)}= 0.  
\end{equation}

When the photon's angular momentum is critical (solid black curve), it orbits the black hole in an unstable circular photon orbit. If the angular momentum is either lower (solid black or green curve) or higher (solid red curve) than this critical value, the photon will either be captured by the black hole or scattered away. In the asymptotic limit, the effective potential affects the trajectory of a light ray coming from a far-off source and moving toward a black hole. The ray may reach a turning point at a certain radius $r_0$ before escaping back toward an observer at infinity. It is well-known that the deflection angle of light increases without bound as it nears the photon sphere. Consequently, this results in the potential for an infinite number of images forming just outside the photon sphere. The unstable photon spheres are characterized by the following conditions \cite{Harko:2009xf}:

We have depicted the behavior of photon orbit radius $r_m$ against the Bumblebee parameter $\ell$ in the left panel of Fig. \ref{plot5}, which shows that $r_m$ decreases with increasing $a$ and $\ell$. The photon orbit radius of the considered black hole can be greater or smaller than the Kerr black hole, depending upon the value of $a$ and $\ell$. We also presented the numerical results of minimum impact parameter $u_m$ versus $\ell$ in the right panel of Fig. \ref{plot5}, which demonstrates that $u_m$ decreases with increasing $a$, whereas it first increases and then decreases as we increase $\ell$.

We can write the angle of deflection for spacetime (\ref{NSR}), as a function of distance of closest approach $r_0$, in following manner \cite{Bozza:2002zj}
\begin{eqnarray}\label{bending1}
\alpha_{D}(r_0)=I(r_0)-\pi,
\end{eqnarray} 
with

\begin{eqnarray}\label{bending2}
I(r_0) = 2 \int_{r_0}^{\infty}\frac{d\phi}{dr} dr
= 2\int_{r_0}^{\infty}\frac{\sqrt{A(r_0) B }\left(2AL+ D\right)}{
\sqrt{4AC+D^2}\sqrt{A(r_0) C-A C(r_0)+L\left(AD(r_0)-A(r_0)D\right)}} dr,
\end{eqnarray}

being the total azimuthal shift in the direction of null geodesics. In the absence of gravitation field (no black hole), $I(r_0)=\pi$, but it grows in the presence of a gravitational field with decreasing $r_0$ such that it exhibits divergence as soon as $r_0$ becomes $r_m$ (unstable photon orbit radius). The analytic solution of the integral (\ref{bending2}) is not possible, hence we followed the approach proposed in Refs. \cite{Bozza:2002zj,Tsukamoto:2016jzh} to get the analytic expression of the deflection angle via expansion of the integral given in Eq. (\ref{bending2}) near $r_m$. Next, we separate the divergent and regular part of $I(r_0)$, by defining a new variable  $z=1 - r_0 / r$ \cite{Ghosh:2020spb,Zhang:2017vap,Tsukamoto:2016jzh}, as follows 
\begin{eqnarray}\label{integral}
I(r_0) &=& I_D(r_0)+I_R(r_0),
\end{eqnarray} with

\begin{eqnarray}
&&I_D(r_0)=\int_{0}^{1} R(0,r_m)f(z,r_0)dz, ~~~~~I_R(r_0)=\int_{0}^{1} [R(z,r_0)f(z,r_0)-R(0,r_m)f(z,r_0)]dz,
\end{eqnarray}

where $R(z,r_0)$ shows regular behaviour for every value of $z$ and $f(z,r_0)$ diverges as $z\to0$.
\begin{figure*}
	\begin{centering}
		\begin{tabular}{ccc}
		    \includegraphics[width=0.5 \textwidth]{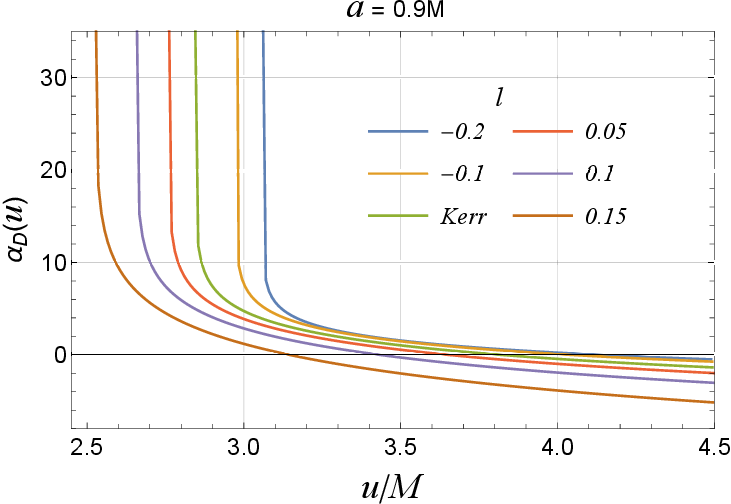}
		    \includegraphics[width=0.5 \textwidth]{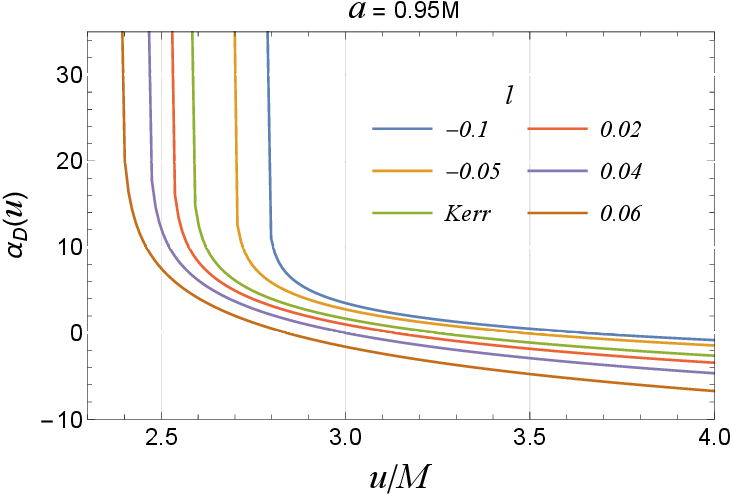}
		    \end{tabular}
	\end{centering}
	\caption{Deflection angle $\alpha_D$ versus impact parameter $u$ for rotating Bumblebee black holes (colored curves) compared to Kerr ($\ell=0$, black). Markers indicate critical impact parameters $u_m$ where $\alpha_D$ diverges, corresponding to photon capture. The logarithmic divergence near $u_m$ (inset) depends on  $\ell$.}\label{plot7}		
\end{figure*} These functions can be given by the following expressions 
\begin{eqnarray}
R(z,r_0)=\frac{2x^2}{r_0} \frac{\sqrt{B}\left(2A(r_0)AL+A(r_0)D\right)}{\sqrt{CA(r_0)}\sqrt{4AC+D^2}},
\end{eqnarray}
\begin{equation}\label{fz}
f(z,r_0)= \frac{1}{\sqrt{A(r_0)-A\frac{C(r_0)}{C}+\frac{L}{C}\left(AD(r_0)-A(r_0)D\right)}}.
\end{equation}
We can approximate the function $f(z,r_0)$  as 
\begin{eqnarray}
f(z,r_0)\sim f_0(z,r_0)=\frac{1}{\sqrt{\eta_1 z +\eta_2 z^2}},
\end{eqnarray}  
where $\eta_1$ and $\eta_2$ can be obtained by using the Taylor expansion of the right-hand side of the above expression. In the limit $r_0\approx r_m$, we can write the deflection angle as \cite{Weinberg:1972kfs,Bozza:2002zj} 

\begin{eqnarray}\label{def}
\alpha_{D}(\theta)=-\bar{a} \log\Big(\frac{\theta D_{OL}}{u_m}-1\Big)+ \bar{b} + \mathcal{O}\left(u-u_m\right),
\end{eqnarray}
where $\theta D_{OL}$ represents the distance of separation of the lens (black hole) from the observer, which is approximately equal to the impact parameter $u$. The strong deflection  coefficients $\bar{a}$ and $\bar{b}$ are defined by the following expressions
\begin{eqnarray}\label{abar}
\bar{a} = \frac{R(0,r_m)}{2\sqrt{{\beta}_m}}, ~~ \bar{b} = -\pi +I_R(r_m) + \bar{a} \log\frac{\gamma r_m^2 }{u_m},
\end{eqnarray} where $\gamma$ can be obtained by using $u-u_m = \gamma (r_0-r_m)^2$. We depicted the numerical results of light deflection angle $\alpha_{D}(\theta)$ in Fig. \ref{plot7} which of course suffers logarithmic divergence when $u\to u_m$. The deflection angle for RBBH at fixed $u<u_m$ is higher than that of the Kerr black hole (black curve in Fig. \ref{plot7}) if the $\ell$ is negative (green curve), and it is less than that of the Kerr black hole if $\ell$ is positive (blue and red curves). The adopted expansion of deflection angle gives a valid approximation for the strong deflection angle only in the neibhourhood of $u_m$. To get a valid approximation of the strong deflection angle in the $u>>u_m$ regime, a different approximation technique will be required.  
 
\subsection{Lensing Observables}
The impact of Bumblebee gravity on the Kerr black holes' strong gravitational lensing observables will be investigated in this subsection. We assume that the source ($S$) and observer ($O$) are sufficiently far enough from the lens (black hole) to be in the flat spacetime. Given the assumption that the source and observer are roughly parallel to the optical axis, the lens equation, which geometrically connects $S$, $O$, and the lens, can be expressed as follows:{Bozza:2018ev,Bozza:2001xd}

\begin{equation}\label{lenseq}
   \beta=\theta-\frac{D_{LS}}{D_{OS}}\Delta \alpha_n,
\end{equation} 
where $\theta$ and $\beta$ represent the angular posision of the source and the image, respectively, with respect to the observer. Furthermore, the distance between the observer and the source is $D_{OS}$, but the distance between the lens and the source is $D_{LS}$. $\alpha_D(\theta)=2n\pi+\Delta \alpha_n$ is an expression for the overall bending angle, $\alpha_D(\theta)$, which incorporates the deflection offset $\Delta \alpha_n$. The number of complete orbits the photon made around the black hole before escaping and reaching the observer is denoted by the positive integer $n$ in this instance. Next, we determined the angular position of the $n$th relativistic picture using Eq. (\ref{def}) in Eq. (\ref{lenseq}) as \cite{Bozza:2002zj}
\begin{equation}
    \theta_n=\theta^0_n+\frac{u_m e_n(\beta-\theta^0_n)D_{OS}}{\bar{a}D_{LS}D_{OL}},
\end{equation} with
\begin{equation}
    e_n=\exp\left(\frac{\bar{b}}{\bar{a}}-\frac{2n\pi}{\bar{a}}\right),~~~~ \text{and}~~~~~\theta^0_n=\frac{u_m(1+e_n)}{D_{OL}}
\end{equation}
\begin{figure*}
	\begin{centering}
		\begin{tabular}{ccc}
		    \includegraphics[width=0.5 \textwidth]{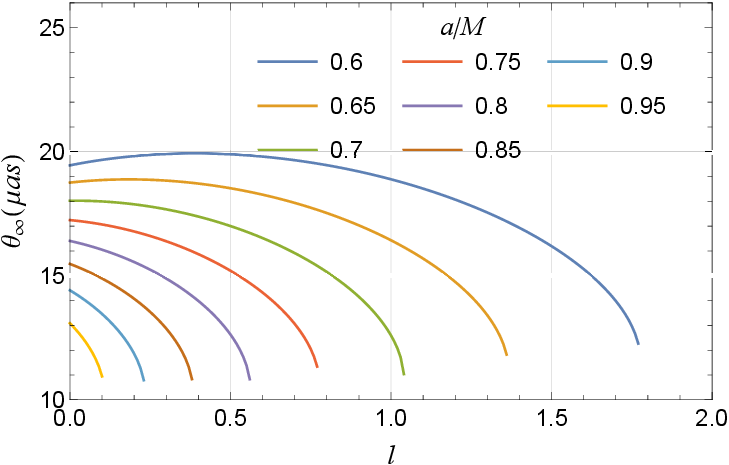}&
		    \includegraphics[width=0.5 \textwidth]{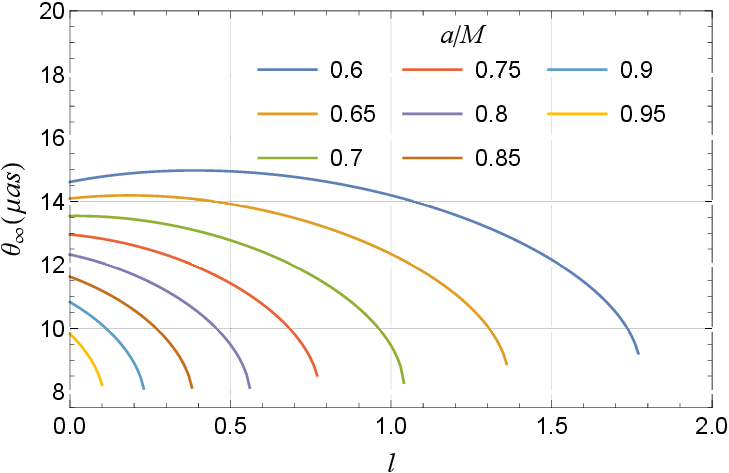}\\
            \includegraphics[width=0.5 \textwidth]{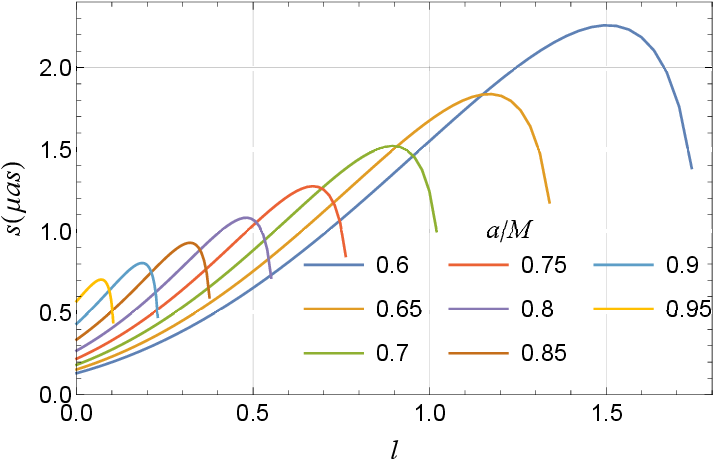}&
		    \includegraphics[width=0.5 \textwidth]{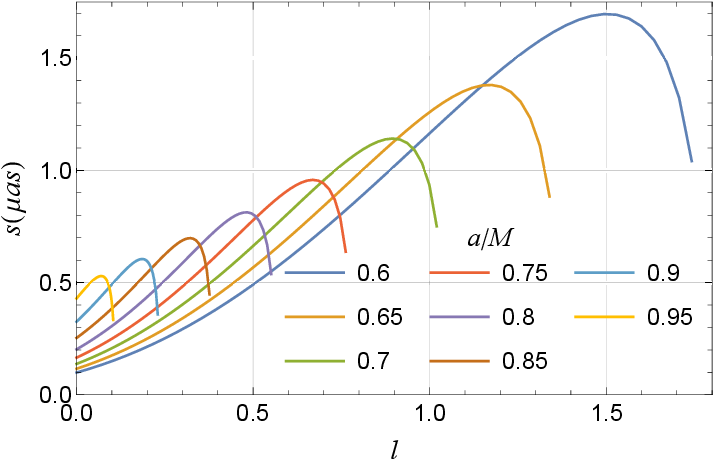}
		    \end{tabular}
	\end{centering}
	\caption{Strong lensing observables for supermassive black holes Sgr\,A* (left) and M87* (right). \textbf{Top:} Angular position $\theta_\infty$ of the photon ring versus Lorentz-violating parameter $\ell$ for different spins $a$. \textbf{Bottom:} Angular separation $s$ between first and higher-order images versus $\ell$ for different spins $a$. }\label{plot8}		
\end{figure*}  
where $\theta^0_n$ is the image angular position when the deflection angle is $2n\pi$. The deflection of the light beam by the gravitational field alters its cross-section, which induces the magnification of the gravitational lensed image. Liouville's theorem states that the deflection of light due to the gravitational field does not affect the surface brightness. Consequently, the ratio of solid angles of the image to the unlensed source can determine the image magnification. Thus, the magnification of $n$th relativistic image can be expressed as \cite{Virbhadra:1999nm,Bozza:2002zj}
\begin{equation}\label{mag}
    \mu_n=\left(\frac{\beta}{\theta}\frac{d\beta}{d\theta}\right)^{-1}\Bigg|_{\theta_n^0}=\frac{u_m e_n(1+e_n)D_{OS}}{\bar{a}\beta D_{LS}D^2_{OL}}.
\end{equation}
We choose the perfect alignment of source with optical axis such that  $\beta\to 0$, Eq. (\ref{mag}) shows divergence, which in another words means that the possibility of detecting an gravitational lensed image is maximum in this case. Also, the magnification of the image reduces exponentially with the order of image $n$, hence, the first image $\theta_1$ is the brightest one. The separation of the outermost image ($n=1$) from the inner packed images ($n=2,3,4,...\infty$) can lead to following observable quantities \cite{Bozza:2002zj}
\begin{eqnarray}\label{observable}
    && \theta_{\infty}=\frac{u_m}{D_{OL}},\\
    && s=\theta_1-\theta_{\infty}\approx \theta_{\infty} \exp\left[\frac{\bar{b}-2\pi}{\bar{a}}\right],\\
    && r_{mag}=\frac{\mu_1}{\sum^{\infty}_{n=2} \mu_{n}}\approx\frac{5\pi}{\bar{a}\log(10)}. 
\end{eqnarray} Here, $\theta_{\infty}$ is the position of packed images, $s$, and $r_{mag}$, respectively, are the angular separation and magnitude difference of fluxes of first and packed images.

Next, we want to move on to another very important strong gravitational lensing observable called the time delay, which possesses information about the time difference between the formation of different relativistic images. As light rays responsible for the formation of different images travel different paths and hence, these light rays reach the observer at different time, i.e., there exists a time lagging between the formation of different images. We can use the procedure given by Bozza and Manchini \cite{Bozza:2003cp} to find the time lag between the formation of different relativistic images. The important requirement to calculate time delay is to consider a light source with varying luminosity; this variation of brightness would manifest in the images with temporal phase depending on the lens geometry. Owing to dimensional variability, time delay can be used to calculate lensing system's mass and length-scale as well as the Hubble parameter in cosmological contexts \cite{1964MNRAS.128..307R,Walsh:1979nx,Blandford:1991xc}. If the first and the second images are on the same side of the lens, the time delay between them can be approximated as \cite{Bozza:2003cp}   
 \begin{equation}
     \Delta T_{21}\approx 2\pi \frac{\tilde{R}(0,r_m)}{\bar{a}\sqrt{\eta_{2_m}}}=\pi u_m,
 \end{equation} where $\tilde{R}(0,r_m)$ is defined as follows
 \begin{equation}
\tilde{R}(z,r_0)= \frac{2x^2\sqrt{B A(r_0)}\left(C-LD\right)}{r_0\sqrt{C(4AC+D^2)}}\left(1-\frac{1}{\sqrt{A(r_0)}f(z,r_0)}\right).
\end{equation}
\begin{figure*}
	\begin{centering}
		\begin{tabular}{ccc}
		    \includegraphics[width=0.5 \textwidth]{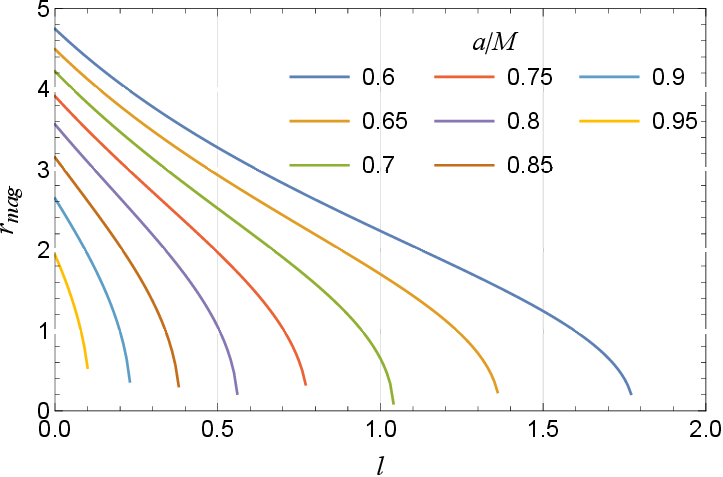}&
    		    \end{tabular}
	\end{centering}
	\caption{Flux magnification ratio $r_{\text{mag}}$ as a function of the Bumblebee parameter $\ell$ for different spin values $a$. The ratio $r_{\text{mag}}$ quantifies the relative brightness of the first relativistic image compared to the others. The curve shows how LSB ($\ell \neq 0$) modifies the image brightness distribution compared to the Kerr case ($\ell = 0$, dashed line). The non-monotonic dependence on $\ell$ suggests competing effects between frame-dragging and Lorentz violation.}\label{plot10}		
\end{figure*}

\subsection{Supermassive black holes, Sgr A* and M87* modeled as RBBH and parameter constraints from EHT observations}
 This subsection is dedicated to analyze the strong gravitational lensing observables of supermassive Sgr A* and M87* black holes modeled as RBBH and comparing these results with Kerr black hole results and finding the constraint on RBBH parameters $a,l$ from the EHT observations. According to most recent observational data \cite {EventHorizonTelescope:2019dse,EventHorizonTelescope:2022wkp}, the mass and distance from earth for Sgr A* are $M\approx 4\times10^6 M_\odot$, and $D_{OL}=8$kpc, whereas for M87* are $M\approx6.5\times10^9 M_\odot$, and $D_{OL}=16.8$Mpc. We depicted the numerical behaviour of lensing observables i.e. $\theta_{\infty}$, $s$ for supermassive black holes Sgr A*, and M87* respectively, in Fig. \ref{plot8} for varying $a$ and $\ell$ from which it can be confirmed that the angular position $\theta_{\infty}$ decreases with increasing $a$ in contrary to separation $s$ which shows increment. When we analysed the behaviour of $\theta_{\infty}$ and $s$ with varying $\ell$, we found that $\theta_{\infty}$ first increases with increasing $\ell$, and it starts decreasing, whereas $s$ shows the same behaviour as it showed with increasing $a$. For negative values of the Bumblebee parameter (green curve), $\theta_{\infty}$ and $s$ are smaller than the Kerr black hole (black curve), but when $\ell>0$ (blue and red curves), these parameters are greater than those for the Kerr black holes. We also plotted the flux magnification parameter $r_{mag}$ vs $\ell$ and $a$ in Fig. \ref{plot10} which shows that $r_{mag}$ decreases for increasing $\ell$ and $a$. The magnification parameter could be greater (green curve) or less (blue and red curves) than that of Kerr black (black holes), depending upon the value of $\ell$ i.e., $\ell<0$ or $\ell>0$, respectively. The numerical results of the lensing observables, $\theta_{\infty}$, $s$, $T_{21}$ and $r_{mag}$ have been tabulated in Table \ref{table1}.

\begin{figure*}
	\begin{centering}
		\begin{tabular}{ccc}
		    \includegraphics[width=0.5 \textwidth]{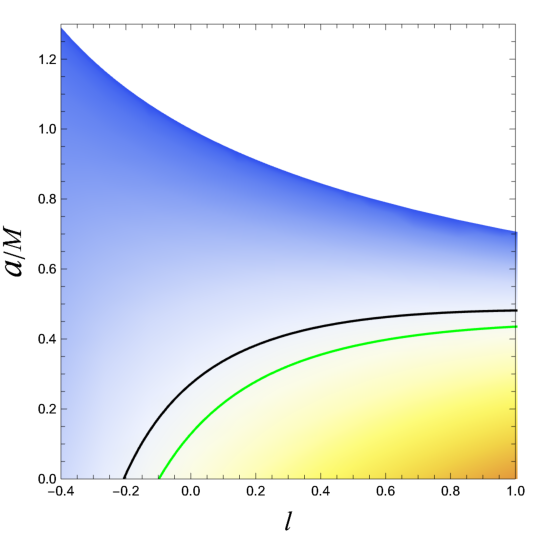}&
		    \includegraphics[width=0.5 \textwidth]{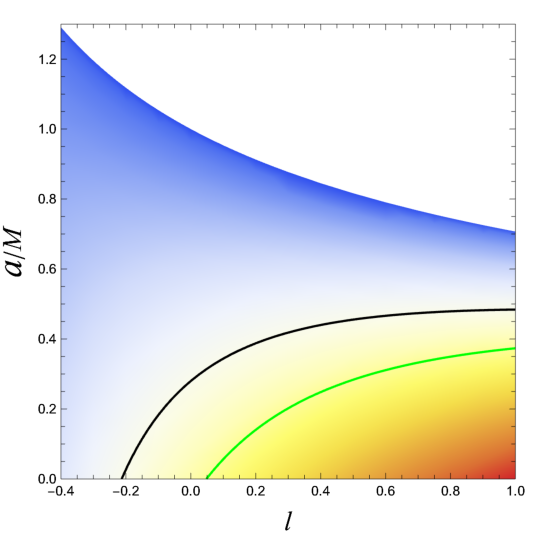}\\
    		    \end{tabular}
	\end{centering}
	\caption{ Shadow angular diameter $\theta_{sh} = 2\theta_{\infty}$ of RBBH as a function of $(a, \ell)$ for Sgr A* (left) and M87* (right). The green and black curves denote $2\theta_{\infty} = 50$ and $2\theta_{\infty} = 46.5$, respectively in the left panel whereas they represent $2\theta_{\infty} = 40.5$ and $2\theta_{\infty} = 37.1$ in the right panel.}   \label{plot11}		
\end{figure*}

The optical detection of black holes is done through a very fascinating geometrical phenomenon called black hole shadows. Black hole shadows are formed due to the winding of photons around the black hole to form the bright photon rings \cite{bardeen1973black,cunningham1973optical}. One of the most effective method to theoretically test the MTG is through constraining the parameters of MTG by using the EHT observations \cite {EventHorizonTelescope:2019dse, EventHorizonTelescope:2022wkp}. To explore the impact of these shadows on the geometry near the horizon, significant analytical and numerical efforts have been devoted to studying and modeling them \cite{Falcke:1999pj,Shen:2005cw,Yumoto:2012kz,Atamurotov:2013sca,Abdujabbarov:2015xqa,Cunha:2018acu,Kumar:2018ple,Afrin:2021ggx,Hioki:2009na, Amarilla:2010zq,Amarilla:2011fx,Amarilla:2013sj,Amir:2017slq,Singh:2017vfr,Mizuno:2018lxz,Allahyari:2019jqz,Papnoi:2014aaa,Kumar:2020hgm,Kumar:2020owy,Ghosh:2020spb,Afrin:2021wlj,Vagnozzi:2022moj,Vagnozzi:2019apd,Afrin:2021imp,Jusufi:2020cpn,Nampalliwar:2021tyz,Jusufi:2022loj,Jafarzade:2023dak}. The EHT has captured shadow images of the supermassive black holes Sgr A* \cite{EventHorizonTelescope:2019dse} and M87* \cite{EventHorizonTelescope:2019dse, EventHorizonTelescope:2022wkp}, with their observed sizes matching Kerr black hole predictions within a $10\%$ margin of accuracy. We shall use the innermost packed images position $\theta_{\infty}$ as the angular size of the black shadow to constrain the parameters $a$ and $\ell$.
\paragraph{Constraints from Sgr A*:}\
{The optical appearance of the supermassive Sgr A* black hole was first released by the EHT collaboration in May 2022, utilizing VLBI technology \cite{EventHorizonTelescope:2022wok,EventHorizonTelescope:2022wkp}. The distance and mass of Sgr A* have been measured using different methods. Observations with the Very Large Telescope Interferometer (VLTI) estimated the distance as $D=9277\pm 9\pm33$ pc} and the mass as $M=(4.297\pm0.013)\times10^6 M_\odot$ \cite{abuter2021improved,GRAVITY:2021xju}. In contrast, data from the Keck Observatory provided estimates of $D=7935\pm50\pm32$ pc for the distance and $M=(3.951\pm0.047)\times10^6 M_\odot$ for mass \cite{Do:2019txf}. The EHT collaboration \cite{EventHorizonTelescope:2022wkp} observed a ring-like structure at a wavelength $1.3$ mm with a diameter $51.8\pm2.3 \mu as$. Based on this, they estimated the mass of Sgr A* to be $M=4.0^{+1.1}_{-0.6}\times10^6 M_\odot$ and the diameter of the black hole shadow to be $\theta_{sh}=(48.7\pm7)\mu as$ assuming a distance $D=8.15\pm0.15$ kpc
\cite{EventHorizonTelescope:2022exc}. The EHT collaboration gave tighter constraints, $46.9\mu as\leq \theta_{sh}\leq 50\mu as$, using different image algorithms such as eht-imaging, SIMLI, and DIFMAP. Using the above EHT data for shadow size, we constrain the RBBH parameters, which are depicted in the right panel of Fig. \ref{plot11}, where the black and green solid curves represent $ \theta_{sh}=46.9 \mu as$ and  $\theta_{sh}=50 \mu as$. It is interesting to note that the RBBH with a lower spin parameter $a$ satisfied the observational data of EHT. For example, for $a=0.1 M$, there exists a constraint on $\ell$ as $-0.096107\leq \ell \leq 0.03894$, and for $a=0.3 M$, the Bumblebee parameter is constrained as $0.12828\leq \ell\leq 0.34926$.
\paragraph{Constraints from M87*:}\
In 2019, the EHT collaboration released the first image of the supermassive black hole M87*, revealing an emission ring with a diameter of $\theta_{er} = 42 \pm 3, \mu as$  \cite{EventHorizonTelescope:2019dse}. Regardless of the parameters $a$ and $\ell$, the Kerr black hole model gives its mass of $M = (6.5 \pm 0.7) \times 10^9 M_\odot$ and distance of $D_{OL} = 16.8$ Mpc and produces the shadow that fits within the 1-$\sigma$ confidence region of the observed data \cite{EventHorizonTelescope:2019dse,EventHorizonTelescope:2019pgp,EventHorizonTelescope:2019ggy}. Accounting for the discrepancy between the emission ring diameter and angular shadow diameter of $\le$10\% \cite{EventHorizonTelescope:2019dse}, the mean angular shadow diameter is $\approx 37.8~\mu as$. The right panel of Fig.~\ref{plot11} shows the angular diameter $\theta_{sh}$ as a function of ($a,\ell$) for RBBH as M87 *, with the green and black curves corresponding to $\theta_{sh}=37.8~\pm 2.7\mu as$. From the right panel of Fig. \ref{plot11}, it can clearly be seen that slowly spinning RBBH satisfy the EHT observations. For instance, for $a=0.1M$ and $a=0.3M$, we have constraint on $\ell$, as $-0.15927\leq \ell\leq 0.14792$ and $0.0273888\leq \ell\leq 0.556945$, respectively.
%The supermassive black hole M87*
\begin{table*}[!htbp]
 \resizebox{\textwidth}{!}{ 
 \begin{centering}
	\begin{tabular}{p{1cm}  p{1.5cm} p{1.6cm} p{1.6cm} p{2.2cm} p{1.5cm} p{1.6cm} p{1.6cm} p{2.2cm} p{1cm} }
\hline\\
\multicolumn{1}{c}{}&
\multicolumn{1}{c}{}&
\multicolumn{3}{c}{Sgr A*}&
\multicolumn{3}{c}{M87*}& \\
{$a$ }&\;\ {$\ell$}  & {$\theta_\infty $ ($\mu$as)} & {$s$ (nas) } & $\Delta T_{2,1}$(min) & {$\theta_\infty $ ($\mu$as)} & {$s$ (nas) } &$\Delta T_{2,1}$(hrs)&\;\; {$r_{mag}$ } 
\\ \hline
\\

\multirow{4}{*}{0.00} &\;-0.30 & 22.0291 & 0.00501304 & 9.61889 & 16.5508 & 0.00376637 & 242.336 &\;\; 8.15371 \\ 
&\; 0.00 & 26.3299 & 0.0329517 & 11.4968 & 19.782 & 0.0247571 & 289.647 &\;\; 6.82188 \\ 
  &\; 0.30 & 30.0207 & 0.109915 & 13.1083 & 22.555 & 0.082581 & 330.249 &\;\; 5.98319 \\
 &\; 0.60 & 33.3049 & 0.259489 & 14.5424 & 25.0225 & 0.194958 & 366.378 & \;\; 5.39317 \\ 
\hline \\
\multirow{4}{*}{0.3} &\;-0.30 & 20.268  & 0.00988279 & 8.8499 & 15.2277 & 0.00742509 & 222.963 &\;\; 7.28835 \\
 &\; 0.00 & 23.1327 & 0.0604049 & 10.1008 & 17.38 & 0.0453831 & 254.477 &\;\; 5.93812 \\
 &\; 0.30 & 25.373  & 0.188768 & 11.079 & 19.0631 & 0.141824 & 279.121 &\;\; 5.08253 \\
 &\; 0.60 & 27.1929 & 0.419499 & 11.8736 & 20.4304 & 0.315176 & 299.141 &\;\; 4.4765 \\ 
\hline \\
\multirow{4}{*}{0.6} &\;-0.30 & 18.2572 & 0.0227698 & 7.9719 & 13.7169 & 0.0171073 & 200.842 & \;\; 6.19966 \\ 
 &\; 0.00 & 19.4504 & 0.131981 & 8.49288 & 14.6133 & 0.0991593 & 213.968 &\;\; 4.74756 \\ 
 &\; 0.30 & 19.9075 & 0.389577 & 8.69249 & 14.9568 & 0.292695 & 218.997 &\;\; 3.78023 \\
 &\; 0.60 & 19.8044 & 0.810663 & 8.64746 & 14.8793 & 0.609064 & 217.862 & \;\; 3.04523\\

\hline
\end{tabular}
\end{centering}}
\caption{The lensing observables $\theta_\infty $ ($\mu$as), $s$ ($\mu$as), time delay $\Delta T_{2,1}$, and $r_{mag}$ for massive black holes Sgr A* and M87.
\label{table1} }
\end{table*}  

\section{Weak Gravitational Lensing}\label{wgl} 
In this section, we focus on studying how light bends, or gravitationally lenses, as it passes near the RBBH, but specifically in the weak gravitational field regime. This means we’re looking at situations where the impact parameter $u$ is much larger than a critical value $u_m$. Under these conditions, the closest distance the light gets to the black hole, $r_0$, is much greater than the radius of the photon orbit, $r_m$. As a result, light rays coming from a distant source reach the observer without making full or multiple loops around the black hole. Instead, they experience only a small bending due to the black hole’s gravity, leading to a deflection angle less than $2\pi$. This setup allows us to examine lensing phenomena that are observable in astrophysical scenarios, where the gravitational influence of the black hole is not excessively dominant. Now, by using Eq. (\ref{coeff}) in Eq. (\ref{bending2}), we obtain the following expression

   \begin{equation}\label{bending3}
    I(r_0) = 2 \int_{r_0}^{\infty}\frac{a(2+\ell r)(-2+r_0)+\beta (-2+r)}{r\left(a^2(1+\ell)+r(-2+r)\right)\sqrt{\frac{(1+\ell)r^3+a^2(2+r+2\ell r)-2a\beta(2+\ell r)-(-2+r)\beta^2}{(1+\ell)r^3}}} dr,
\end{equation}
with 
\begin{equation}
    \beta=\frac{-a(2+\ell r_0)+(1+\ell)r_0\sqrt{a^2+\frac{(-2+r_0)r_0}{1+\ell}}}{-1+r_0}
\end{equation}
\begin{figure*}
	\begin{centering}
		\begin{tabular}{ccc}
		    \includegraphics[width=0.5 \textwidth]{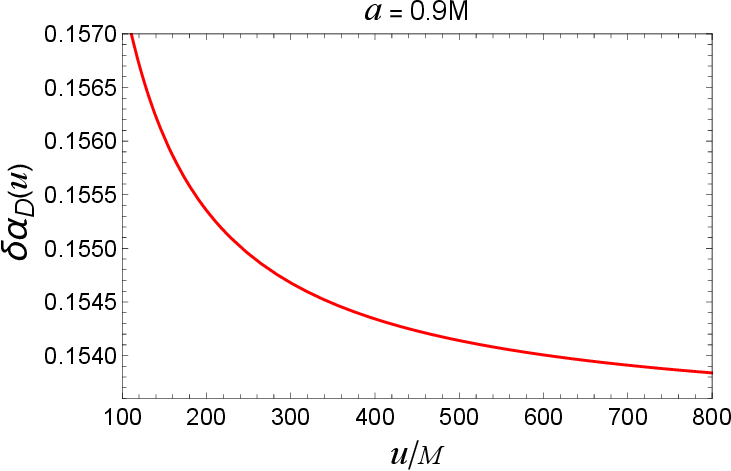}&
		    \includegraphics[width=0.5 \textwidth]{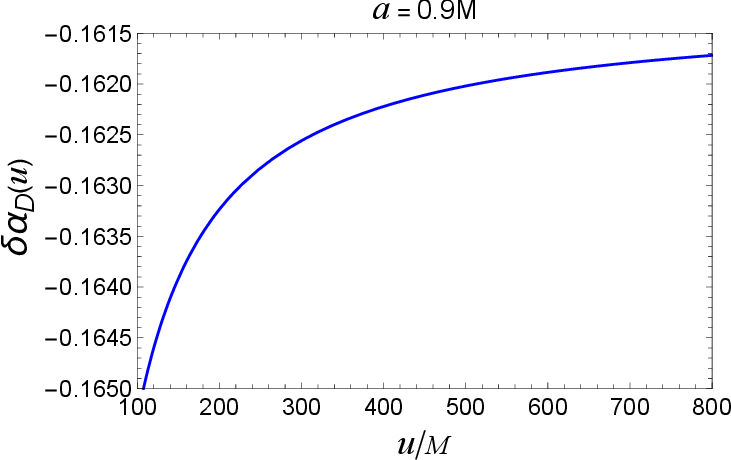}&
    		    \end{tabular}
	\end{centering}
	\caption{The deviation in weak lensing deflection angle $\delta\alpha_{D}(u)=\alpha_D(u)-\alpha_D(u)|_{Kerr}$ vs impact parameter $u$ of RBBH from Kerr black holes. The red and blue curves are plotted for $\ell=0.1$ and $\ell=-0.1$, respectively.}\label{plot12}		
\end{figure*}
\begin{figure*}
	\begin{centering}
		    \includegraphics[width=0.5 \textwidth]{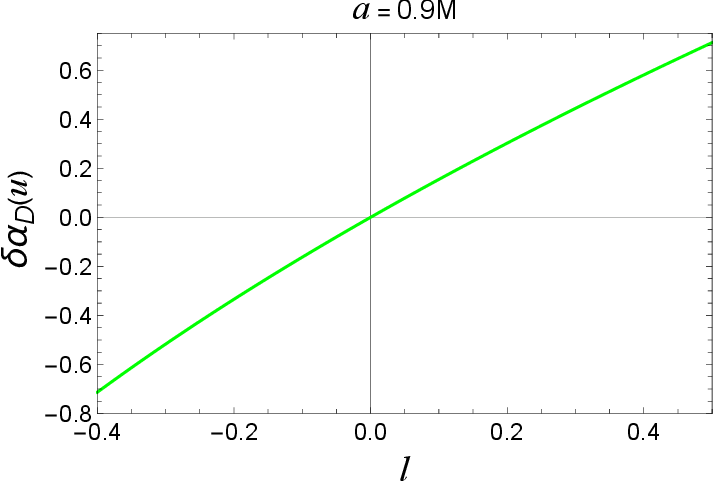}	   
	\end{centering}
	\caption{The deviation in weak lensing deflection angle $\delta\alpha_D(u)=\alpha_D(u)-\alpha_D(u)|_{Kerr}$ vs  $\ell$ of RBBH from Kerr black holes. We kept $u=300$ for this graph.}\label{plot13a}		
\end{figure*} 
As we consider the weak gravitational field regime, we define a new variable $y=\frac{r_0}{x}$ and take the Taylor expansion of the integrand in Eq. (\ref{bending3}) to get 

    \begin{eqnarray}
       &&I(r_0)\approx \int_{0}^{1}2\sqrt{\frac{1+\ell}{1-y^2}}+\frac{2(1+y+y^2)\sqrt{1+\ell}}{r_0(1+y)\sqrt{1-y^2}}-\frac{\sqrt{1+\ell}\left[12(1+y+y^2)-4a\sqrt{1+l}(1+y)\left(4-a\sqrt{1+l}(1+y)(-1+2y^2)\right)\right]}{4(1+y)^2\sqrt{1-y^2}r_0^2}\nonumber\\
       &&-\frac{(1+\ell)}{8\sqrt{1-y^2}(1+y)^3r_0^3}
       \Bigg[-\frac{5(1+y+y^2)^3}{\sqrt{1+\ell}}+8a(2+6y+9y^2+5y^3)+12a^2\sqrt{1+\ell}(1+y)^2\left(1-y-y^2(1+2y+2y^2)\right)\Bigg]\nonumber\\&&+\mathcal{O}\left(\frac{1}{r_0^4}\right)
    \end{eqnarray}
    By integrating above integrand and using Eq. (\ref{bending1}), we obtain the value of deflection angle as
   
        \begin{eqnarray}\label{bending4}
       \alpha(r_0)&=&\pi(-1+\sqrt{1+\ell})+\frac{4\sqrt{1+\ell}}{r_0}+\frac{1}{r_0^2}\Bigg(\sqrt{1+\ell}\left(-4+\frac{15\pi}{4}\right)-4a(1+\ell)\Bigg)\nonumber\\&&+\frac{1}{r_0^3}\Bigg(\sqrt{1+\ell}\left(-\frac{15\pi}{2}+\frac{122}{3}\right)+2a(1+\ell)\left(8-5\pi+a\sqrt{1+\ell}\right)\Bigg)+\mathcal{O}\left(\frac{1}{r_0^4}\right).
   \end{eqnarray} 
   The above expression of deflection angle clearly shows that it decreases with increasing closest approach distance $r_0$. Next, we want to get the deflection angle in terms of impact parameter $u$, for which we firstly expanded $1/r_0$ in a series of $1/u$ as follows 
   
       \begin{eqnarray}\label{expansion}
     \frac{1}{r_0}=\frac{\sqrt{1+\ell}}{u}+\frac{1+\ell-2a\ell\sqrt{1+\ell}}{2u^2}+\frac{5(1+\ell)^{3/2} +4a(1+2\ell+3\ell^2)\left(a\sqrt{1+\ell}-2\right)}{8u^3}+\mathcal{O}\left(\frac{1}{u^4}\right)  
   \end{eqnarray} 
   By inserting Eq. (\ref{expansion}) in Eq. (\ref{bending4}), we derive the deflection angle in terms of the impact parameter as
        \begin{eqnarray}\label{bending5}
       \alpha(b)&=&(-1+\sqrt{1+\ell})\pi+\frac{4(1+\ell)M}{b}+\frac{(1+\ell)M}{4b^2}\left(\frac{15\pi\sqrt{1+\ell} M}{16}-16a(1+2\ell)\right)+\frac{(1+\ell)M}{6b^3}\Bigg(128(1+\ell)M^2\nonumber\\&&-15\pi \sqrt{1+\ell}(4+7\ell)aM+24a^2(1+4\ell M)^2\Bigg)+\mathcal{O}\left(\frac{1}{b^4}\right).
   \end{eqnarray} Here, $b=uM$ is signifying the rescaled impact parameter. In the limiting case $\ell\to0$, Eq. (\ref{bending5}) reduces to the deflection angle for Kerr black holes \cite{Ovgun:2018fte,Li:2020zxi,He:2021rrz}. We have plotted the deviation of weak gravitational deflection angle of RBBH from Kerr black holes in Fig. \ref{plot11} from which it can be confirmed that for positive value of $\ell$ (red curve), RBBH have larger deflection than Kerr black holes whereas for $\ell<0$ (blue curve), the deflection for RBBH is smaller than Kerr black holes. Fig. \ref{plot12} also shows that the deviation is more significant for lower values of $u$. In the region $\ell<0$, the magnitude of correction in deflection angle decreases for increasing $\ell$ whereas when $\ell>0$, the correction increases monotonically with $\ell$ (cf. Fig. \ref{plot13a})   
   
\subsection{The size of the Einstein ring }  
Next, we further calculate the analytical expression of the angular radius of Einstein’s ring. We merely consider the extraordinary situation that the source, lens, and observer are aligned along the same axis, and the source and observer are located in an asymptotic region. In addition, we estimate the upper bound of the Lorentz-violating coefficient $\ell$ by using observational data of Einstein's ring of the galaxy ESO325-G004.

Without assuming the intersection point between the incoming and outgoing ray trajectories lies on the lens plane, Bozza derived an improved version of the lens equation as~\cite{Bozza:2008ev}
\begin{align}
d_S\tan\mathcal{B}=\dfrac{d_L\sin\vartheta-d_{LS}\sin(\hat{\alpha}-\vartheta)}{\cos(\hat{\alpha}-\vartheta)},\label{Bozzalensequation}
\end{align}
From the figure, $\mathcal{B}$ is the angular position of the unlensed source; $\vartheta$ is the angular position of an image;  $d_L$ is the angular diameter distance from the observer to the lens plane; $d_{LS}$ is the angular diameter distance from the lens plane to the source plane; $d_S=d_{LS}+d_L$ and $\sin\vartheta=b/d_L$.

The source, lens, and observer are aligned that leads to $\mathcal{B}=0$, and thus in the weak deflection approximation. It is clearly seen that $\hat \alpha$ and $\vartheta$ are of the order $\mathcal{O}(\varepsilon)$ as $\mathcal{B}=0$ from (\ref{bending5}) and (\ref{Bozzalensequation}). the (\ref{Bozzalensequation}) is written as
\begin{align}
\vartheta_E= \frac{d_{LS}}{d_S}\hat{\alpha}+\mathcal{O}\left(\varepsilon^3\right),\label{Einsteinring}
\end{align}
where $\vartheta_E$ denotes the angular radius of Einstein's ring.
Putting (\ref{bending5}) and
\begin{align}
\vartheta_E = \arcsin\left(\dfrac{b}{d_L}\right) = \dfrac{b}{d_L} + \mathcal{O}\left(\varepsilon^3\right)
\end{align}
into (\ref{Einsteinring}), we obtain
\begin{align}
\vartheta_E =\dfrac{d_{LS}}{d_S}\left[\dfrac{4M}{d_L\vartheta_E}+\dfrac{3\pi \ell}{8}+\dfrac{15\pi}{4}\dfrac{M^2}{d_L^2\vartheta_E^2}+\dfrac{3\pi \ell^2}{128}+2\ell\dfrac{M}{d_L\vartheta_E}\right]+ \mathcal{O}\left(\varepsilon^3\right).\label{new-vartheta2}
\end{align}
\begin{figure}
	\begin{centering}
		    \includegraphics[width=0.5 \textwidth]{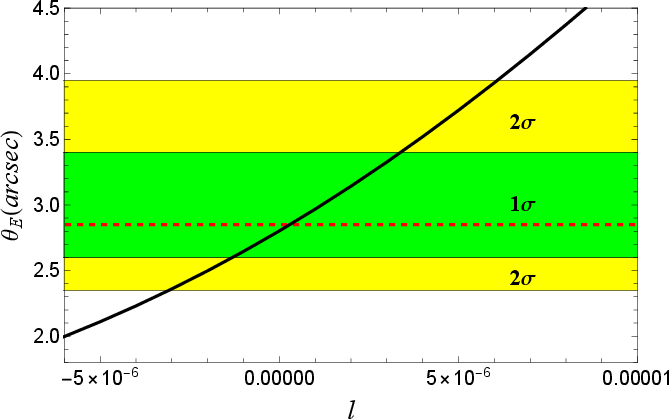}
    		   
	\end{centering}
	\caption{The estimation of angular radius of Einstein ring $\theta_E$ vs the Bumblebee gravity parameter $\ell$ with the Hubble Space Telescope observations. The uncertainties in $1\sigma$ and $2\sigma$ confidence levels are represented by green and yellow shaded regions, respectively.}\label{plot13}		
\end{figure}
Taking the galaxy ESO325-G004 as the lens, our aim here is to estimate the upper bound of $\ell$ from the observational data of the Einstein ring. All the data below are from \cite{Smith:2005pq,Smith:2013ena}, and we use $G\neq 1$ and $c\neq 1$ in this sector analysis and calculation. The observed value of the Einstein ring by lensing by the elliptical galaxy ESO325-G004 at $z_l=0.0345$ is
\begin{align}
\vartheta_E^{obs}=(2.85_{-0.25}^{+0.55})''.\label{obsering}
\end{align}
The mass of the galaxy ESO325-G004 is $M=1.50\times 10^{11} M_{\odot}$, where $M_{\odot}=1.98\times 10^{30}{\rm kg}$ is the mass of the Sun. Here, the total mass $M$ inside Einstein's ring of the galaxy ESO325-G004 not only includes black hole mass, but also incorporates the mass of dark matter and luminous matter.. The redshift of the background source galaxy $z_s=1.141$. Hubble has found that the relation between the spectrum redshift and distance of galaxy i.e. $cz=H_0 D$, where $H_0=70.4\, {\rm km}\, {\rm s}^{-1}\,({\rm Mpc})^{-1}$ is the Hubble constant and $D$ denotes the proper distance, related to the comoving distance $d$ by $d=D(1+z)$ for a flat cosmology. Then we have
\begin{align}
d_S=\dfrac{c z_s(1+z_s)}{H_0}=2.863\times10^4\,{\rm Mpc}, \quad d_L=\dfrac{c z_l(1+z_\ell)}{H_0}=1.542\times10^2\,{\rm Mpc} \ .\label{ESOdata}
\end{align}

According to the above data, the angular radius of the Einstein ring (\ref{new-vartheta2}) as a function of the $\ell$ is shown in Fig. \ref{plot13}. It is found that the $\vartheta_E$ increases gradually with the LSB coefficient $\ell$ increases. From Fig. \ref{plot13}, we clearly see that $\ell$ can be constrained as $\ell\lesssim 4.472\times 10^{-6}$ within $1\sigma$ uncertainties and $\ell\lesssim 8.084\times 10^{-6}$ within $2\sigma$ uncertainties. In addition, by using the average error formula $(\sum|\ell_i-\bar{\ell}|/n)$, we do every rough with error analysis, where $\ell_i$ denotes every measured value, $\bar{\ell}$ represents the average value, and $n$ is the number of measured value. Finally, the average error with $\ell\simeq4.472\times 10^{-6}$ and $8.084\times 10^{-6}$ is about $1.806\times 10^{-6}$.

\section{Conclusions}\label{conclusion}
From the precession of Mercury's perihelion to the most recent imaging of black hole shadows by the EHT, GR has withstood a century of observational testing. However, GR is anticipated to fail when quantum gravitational effects become important at high energy scales. Moreover, it assumes Lorentz invariance (LI) as a fundamental symmetry of spacetime. Yet, various approaches to quantum gravity—including string theory and loop quantum gravity—suggest that this symmetry may be violated at Planckian scales. Here, spontaneous Lorentz symmetry breaking through a vector field, obtaining a nonzero vacuum expectation value, is incorporated into an efficient field theory framework offered by the Bumblebee gravity model.  Exploring black hole solutions in such modified gravity theories is theoretically appealing and observationally timely, as current and upcoming high-resolution astronomical facilities offer an unprecedented opportunity to test deviations from GR through gravitational lensing, shadow structures, and relativistic image formation around black holes. 

The principle of LI is fundamental to our understanding of spacetime. It serves as a cornerstone for both GR and the Standard Model (SM) of particle physics—two highly successful field theories describing the universe. GR explains gravitation in the classical framework, while SM addresses particles and the other three fundamental interactions within a quantum context. However, LI may not hold as an exact symmetry at all energy scales, especially when quantum gravity effects are taken into account \cite{Mattingly:2005re}. Hence, the Bumblebee model got attention, which shows LSB, and black hole solutions are also discussed \cite{Casana:2017jkc,Ding:2020kfr,Ding:2019mal,An:2024fzf}. Building on this, we obtained a rotating part of the black hole solution obtained \cite{Casana:2017jkc} and explored the strong gravitational lensing of these black holes. 

We first obtained the RBBH by applying a modified Newman-Janis algorithm to a spherically symmetric static solution incorporating spontaneous LSB. The RBBH metric is characterized by the LSB parameter $\ell$ apart from the mass $M$ and spin parameter $a$. Significant deviations from the Kerr geometry are caused by the parameter $\ell$, providing a way to investigate Lorentz symmetry violation in strong gravity regimes. The black hole's horizon structure and photon dynamics are drastically altered by the Bumblebee gravity framework, which alters the classical Kerr geometry. After analyzing the horizon topology and causal structure of the RBBH spacetime, we discovered that $\ell$ has a nontrivial influence on the existence requirements and the extremal bound. In particular, the event horizon radius deviates from the Kerr case for fixed $a$, increasing with negative $\ell$ and decreasing with positive $\ell$.

We then investigated null geodesics and computed the strong deflection angle $\alpha_D$ via Bozza's formalism. The study of photon dynamics unveiled that the unstable photon orbit radius $r_m$ and the minimum impact parameter $u_m$ decrease with increasing $a$ and $\ell$. Consequently, it influences the deflection angle $\alpha_D$, which varies based on $\ell$ but displays the anticipated divergence close to $u_m$. In particular, the deflection angle is smaller than Kerr for $\ell > 0$ and larger for $\ell < 0$, demonstrating how Lorentz-violating effects appear in observation.

Using the strong deflection limit formalism, we derived the key lensing observables—such as the angular position of relativistic images $\theta_\infty$, the angular separation $s$, and the flux ratio $r_{\mathrm{mag}}$—for RBBHs and contrasted them with those for Kerr black holes. We modelled the lensing by supermassive black holes Sgr A* and M87*, incorporating their mass and distance values consistent with EHT observations. Strong lensing observables—such as the angular image position $\theta_\infty$, angular separation $s$, and relative flux ratio $r_{\text{mag}}$—were computed and shown to vary sensitively with both $\ell$ and $a$. The time delay $\Delta T_{21}$ between relativistic images also varies significantly, with measurable deviations for both Sgr A* and M87*. Significantly, for appropriate parameter choices, the observables lie within the $1\sigma$ bounds of EHT measurements, constraining $\ell$ to specific subspaces of the $(a, \ell)$ parameter space. Depending on the parameters of $a$ and $\ell$, the time delay between the first and second relativistic pictures was also analysed, producing results of almost 370 hours for M87* and roughly 15 minutes for Sgr A*. These parameters offer further discriminators between Kerr and non-Kerr geometries, despite their difficulty in measurement.

Importantly, our analysis demonstrated that the predicted ranges of $\theta_\infty$ for RBBH can lie within the $1\sigma$ bounds of the angular diameter measurements reported by the EHT for Sgr A* and M87*, thus placing constraints on the allowed parameter space for $(a, \ell)$. 
Finally, weak lensing analysis and Einstein ring size comparison with HST data from galaxy ESO325-G004 yielded an upper bound of $\ell \lesssim \mathcal{O}(10^{-6})$, reinforcing the observational viability of the RBBH solution.

Our research demonstrates that Lorentz symmetry violation in Bumblebee gravity has theoretically consistent effects that may be observed by strong gravitational lensing. Such departures from GR may be crucially constrained by future high-resolution astronomical observations, including those from the next-generation EHT, which will provide a new window into quantum gravitational effects and basic symmetry breaking in nature.

In conclusion, the RBBH solution provides phenomenologically rich and observationally distinguishable deviations from the Kerr black hole in both strong and weak lensing regimes. Our findings confirm that gravitational lensing—particularly when used in conjunction with high-precision imaging using instruments like the EHT—is a potent probe of Lorentz symmetry violation and modified gravity in the strong-field regime. Future high-resolution observations of ngEHT \cite{Ayzenberg:2023hfw} may further narrow the bounds on the Bumblebee parameter and separate RBBH spacetimes from standard GR black holes. Further, extrapolation of this analysis to black hole shadows, polarimetric signatures, and gravitational wave echoes may offer complementary probes. Perturbative dynamics and quasinormal mode investigations in the background RBBH also form a natural line of future study.

\section{Acknowledgments} 
S.G.G. gratefully acknowledges the support from ANRF through project No. CRG/2021/005771. This work is supported by the Zhejiang Provincial Natural Science Foundation of China under Grants No.~LR21A050001 and No.~LY20A050002, the National Natural Science Foundation of China under Grants No.~12275238 and No. ~W2433018, the National Key Research and Development Program of China under Grant No. 2020YFC2201503, and the Fundamental Research Funds for the Provincial Universities of Zhejiang in China under Grant No.~RF-A2019015. A.K. would like to express his gratitude to the ITPC members for their warm hospitality.

\bibliographystyle{apsrev4-1}
\bibliography{Bumblebee.bib}
\end{document}